\documentclass[12pt]{iopart}
\usepackage{amsfonts}

\newcommand{\bV}{\mbox{\boldmath{$V$}}}

\newcommand{\bv}{\mbox{\boldmath{$v$}}}

\newcommand{\bu}{\mbox{\boldmath{$u$}}}

\newcommand{\bJ}{\mbox{\boldmath{$J$}}}

\newcommand{\mR}{\mathbb{R}}
\newcommand{\mN}{\mathbb{N}}

\usepackage{color}
\usepackage{graphicx}
\begin{document}

\title[First eigenvalue/eigenvector in sparse random matrices]{First eigenvalue/eigenvector in sparse random symmetric matrices: influences of degree fluctuation}

\author{$^\dag$Yoshiyuki Kabashima and $^\ddag$Hisanao Takahashi}

\address{$^\dag$Department of Computational Intelligence and Systems Science, 
Tokyo Institute of Technology, 
Yokohama 226-8502, Japan \\
$^\ddag$The Institute of Statistical Mathematics, 10-3 Midori-cho, Tachikawa, Tokyo 190-8562, Japan}
\ead{$^\dag$kaba@dis.titech.ac.jp, $^\ddag$hisanao@ism.ac.jp}

\begin{abstract}
The properties of the first (largest) eigenvalue and its eigenvector (first eigenvector) are investigated for large sparse random symmetric matrices that are characterized by bimodal degree distributions. In principle, one should be able to accurately calculate them by solving a functional equation concerning auxiliary fields which come out in an analysis based on replica/cavity methods. However, the difficulty in analytically solving this equation makes an accurate calculation infeasible in practice. To overcome this problem, we develop approximation schemes on the basis of two exceptionally solvable examples. The schemes are reasonably consistent with numerical experiments when the statistical bias of positive matrix entries is sufficiently large, and they qualitatively explain why considerably large finite size effects of the first eigenvalue can be observed when the bias is relatively small.

\end{abstract}

\maketitle

\section{Introduction}
Since their introduction by Wigner for approximating the complex Hamiltonian of heavy nuclei, random matrices have been used in various fields of physics and other disciplines. The list of applications includes nuclear theory \cite{Wigner}, quantum chaos \cite{QuantumChaos}, localization in electron systems \cite{Localization}, finance \cite{Finance}, complex networks \cite{ComplexNetworks}, wireless communication \cite{TurinoVerdu}, combinatorial problems in computer science \cite{Alon1997}, and more. 

In general, the purpose of random matrix theory (RMT) is to investigate the statistical properties of physical quantities that are defined by samples drawn from an ensemble of $N\times N$ random matrices. A major topic is the evaluation of the asymptotic eigenvalue spectrum, which is the typical distribution of eigenvalues as $N \to \infty$. For a Gaussian orthogonal ensemble (GOE), whose matrix entries are independently distributed obeying identical Gaussian distributions of zero mean, the spectrum follows the Wigner semicircle distribution \cite{Mehta1967}. Another type of asymptotic spectrum, termed the Mar\u{c}enko-Pastur distribution, comes from the covariance matrix of rectangular matrices whose entries are independently sampled from identical Gaussian distributions of zero mean \cite{MarchenkoPastur1967}. Recent developments on sparsely connected disordered systems have led to significant progress in being able to analyze the spectrum of sparse random matrices \cite{BrayRodgers1988,BiroliMonasson1999,SemarjianCugliandolo2002,NagaoTanaka2007,Nagao2008,Rodgers2008,Kuhn2008,MetzNeriBolle2010,Rodgers2010,Bordenave2010,Jort2011,MetzNeriBolle2011}.

Evaluation of the first (largest) eigenvalue and its eigenvector (first eigenvector) is another major topic of RMT. For a GOE, the first eigenvalue converges to $2$ when the variance of the matrix entries 
is provided
as $N^{-1}$ in the limit of $N \to \infty$, and the finite size correction follows the Tracy-Widom distribution when $N$ is large but finite \cite{TracyWidom1993}. As for the covariance matrix of dense random rectangular matrices, the asymptotic behaviors of the first eigenvalue/eigenvector have been examined analytically and numerically in situations where one can set the strength of preferential directions underlying the random rectangular matrices \cite{HoyleRattray2004,HoyleRattay2006}. 
For sparse matrices, convergence to the Tracy-Widom distribution was recently proved for the first eigenvalue 
in the case of fixed degrees, which denote the numbers of nonzero entries per row/column in matrices, and entries of random signs \cite{Sodin2009}. 
There are also various studies on the second eigenvalue 
of adjacency matrices of fixed degrees \cite{Miller2008,Friedman2004}. 
However, as far as the authors know, 
the first eigenvalue problem for ensembles of sparse matrices has not been sufficiently examined yet, despite there being analyses of their spectrum. Moreover, the need for an accurate solution to the first eigenvalue problem seems to be growing, because the first eigenvector is useful for extracting valuable information in the field of data analysis \cite{PageRank2006,Recommendation2008} and in constructing approximate solutions of various combinatorial problems \cite{Alon1997,CS}. 

In light of this potential need, we herein investigate the asymptotic properties of the first eigenvalue/eigenvector for ensembles of sparse symmetric matrices. A preliminary investigation indicated that the properties are considerably influenced by the {\em fluctuation of degrees} \cite{JPCS2010}.
 In order to be able to control the influence of the degree fluctuation and the properties of the first eigenvector in a simple manner, we focus on matrix ensembles that are characterized by a bimodal degree distribution and a biased binary distribution of nonzero matrix entries. 

This article is organized as follows. In the next section, we explain the model that we will examine. Section \ref{section3} introduces the methodological bases for analyzing the model. Although the model we investigate seems quite simple, analyzing it exactly in general situations is technically difficult. Nevertheless, one can still analytically solve the problem in two specific cases, which are shown in section \ref{section4}. On the basis of lessons derived from the solvable cases, we develop approximate assessment schemes for handling a more general situation in section \ref{section5}. The final section is devoted to a summary. 

Some contents in the following are shared with a conference paper \cite{JPCS2010}.
Precisely, the model to examine is identical, and 
the methodology in section 3.2 and the solvable example in section 4.1 were
shown for the first time in the conference paper.  
The other parts are, however, newly provided in the present article.  

\section{Model definition}
\label{section2}
We consider a sparse network of $N$ nodes indexed by $i=1,2,\ldots,N$. The network is characterized by a bimodal distribution $p(k)=p_1\delta_{k,c_1} + p_2 \delta_{k,c_2}$ of degree $k(=0,1,2,\ldots)$, which stands for the number of links from each node to other nodes. Here, we assume that $p_1,p_2 \in [0,1]$ satisfy $p_1+p_2=1$ and $c_1 \le c_2 $. Moreover, $\delta_{x,y}=1$ if $x=y$, and it vanishes, otherwise. We denote the average degree as $\overline{c}=p_1c_1+p_2c_2$, and suppose that the network is constructed randomly for aspects other than degree. A practical scheme for generating such a network is basically as follows \cite{StegerWormald1999}:
\begin{quote}
\item[(S)] Set $k_i=c_1$ and $k_i=c_2$ for $Np_1$ and $Np_2$ indices of $i=1,2,\ldots,N$, respectively, and make a set of indices $U$ to which each index $i$ attends $k_i$ times. Accordingly, steps (A)-(C) are iterated as follows. 
\item[(A)] Choose a pair of two different elements from $U$ randomly. 
\item[(B)] Denote the values of the two elements as $i$ and $j$. If $i \ne j$ and the pair of $i$ and $j$ have not been chosen up to that moment, make a link to the pair, and remove the two elements from $U$. Otherwise, return them back to $U$. 
\item[(C)] If $U$ becomes empty, finish the iteration. Otherwise, if there is no possibility that any more links can be made by (A) and (B), return to (S). 
\end{quote}

Once we have generated the network, we assign entries $J_{ij}=\pm 1$ to the links generated by the above procedure, 
sampling $\pm 1$ randomly and independently from a biased binary distribution: 
\begin{eqnarray}
p_J(J_{ij}|\Delta)=\frac{1+\Delta}{2} \delta_{J_{ij},1}+\frac{1-\Delta}{2} \delta_{J_{ij},-1}. 
\label{biased_dist}
\end{eqnarray}
We set $J_{ij}=0$ if $i$ and $j$ are not connected in the network. This yields a sample $\bJ=(J_{ij})$ of sparse random symmetric matrices that we will focus on. 

The objective of our study is to investigate how the properties of the first eigenvalue $\Lambda$/eigenvector $\bV=(V_1,V_2,\ldots,V_N)$ of the random matrix $\bJ$ depend on the system parameters $c_1, c_2, p_1=1-p_2$ and $\Delta$ as $N$ tends to infinity. A simple consideration guarantees that $\Lambda$ is upper bounded by $c_2$ for any realization of $\bJ$ (\ref{LambdaUB}). On the other hand, when $p_1=1$ and $\Delta=1$, which means each row/column of $\bJ$ has $c_1$ entries of unity exactly, $\Lambda=c_1$ and $\bV \propto (1,1,\ldots,1)^{\rm T}$ hold, where ${\rm T}$ denotes the matrix transpose operation. This implies that the inverse participation ratio (IPR) of $\bV$, ${\rm IPR}\equiv (\sum_{i=1}^N V_i^4)/(\sum_{i=1}^N V_i^2)^2$, converges to zero, and therefore $\bV$ extends over almost all nodes as $N$ tends to infinity in the vicinity of this parameter setting. However, earlier studies have indicated that $\bV$ can be localized in the vicinity of a few nodes being characterized by a finite IPR  when a small number of nodes of larger degree $c_2$ are added to the sparse network and if $c_2$ is sufficiently large \cite{BiroliMonasson1999,SemarjianCugliandolo2002,MetzNeriBolle2010}. One of our interests is to clarify how such a change in the profile of $\bV$ is related to the value of $\Lambda$.

\section{Analytical bases: replica and cavity methods}
\label{section3}
\subsection{Replica method}
Formulating the first eigenvalue problem as 
\begin{eqnarray}
\Lambda=\frac{1}{N} \mathop{\rm max}_{\bv}\left \{
\bv^{\rm T} \bJ \bv\right \}
\ \mbox{ subj. to } \ |\bv|^2=N, 
\label{eq:quadopt}
\end{eqnarray}
will form the basis of our analysis. Here, $\mathop{\rm max}_X\{ f(X) \}$ denotes maximization of a function $f(X)$ with respect to $X$. The solution to this problem accords with $\bV$. Identifying 
$-(1/2) \bv^{\rm T} \bJ \bv$ as the Hamiltonian of the dynamical variable $\bv$ yields the partition function, 
\begin{eqnarray}
Z(\beta; \bJ)=\int d \bv \delta \left (|\bv|^2 -N \right )\exp \left (\frac{\beta \bv^{\rm T} \bJ \bv}{2} \right ). 
\label{eq:partition}
\end{eqnarray}
This offers another way to express the first eigenvalue: $\Lambda=2\lim_{\beta \to \infty} (N\beta)^{-1}\ln Z(\beta;\bJ)$. 
The typical first eigenvalue can be obtained by averaging the logarithm of the partition function over the random matrix $\bJ$.

The above considerations naturally lead one to consider trying a solution using the replica method \cite{replica}. Hereafter, let us generally denote $[O(X)]_X$ as the average of $O(X)$ with respect to random variable $X$. In the replica method, we first evaluate analytical expressions of the moment of $Z(\beta;\bJ)$, $\left [Z^n(\beta;\bJ) \right]_{\bJ}$, for $\forall{n}=1,2,\ldots \in \mN$ utilizing an identity $Z^n(\beta;\bJ)=\int \left (\prod_{a=1}^n d\bv^a \delta(|\bv^a|^2-N) \right ) \times \exp \left ((\beta/2) \sum_{a=1}^n (\bv^a)^{\rm T} \bJ \bv^a \right )$, which is valid for only $n \in \mN$. The integration variables $\bv^a$ $(a=1,2,\ldots,n)$ are sometimes termed ``replicas'' since they can be regarded as $n$ copies of the original variable $\bv$ that share the identical external random coupling $\bJ$. Although the identity is valid for only $n \in \mN$, the expressions of $\left [Z^n(\beta;\bJ) \right]_{\bJ}$ evaluated with the saddle point method for $N \gg 1$ under appropriate assumptions about the permutation symmetry of the replica indices $a=1,2,\ldots,n$ are likely to hold for $n \in \mR$ as well. Therefore, we can employ the analytical expressions for computing the average of the logarithm of the partition function by utilizing the identity $N^{-1} \left [\ln Z(\beta;\bJ) \right]_{\bJ}=\lim_{n \to 0} (\partial/\partial n) N^{-1} \ln \left [Z^n(\beta;\bJ) \right]_{\bJ}$. In particular, under the replica symmetric (RS) ansatz, which implies that the saddle point is invariant under any permutation of the replica indices, this yields an expression for the typical first eigenvalue as
\begin{eqnarray}
[\Lambda]_{\bJ}=\mathop{\rm extr}_{q(\cdot),\widehat{q}(\cdot), \lambda}\left \{
\frac{\overline{c}}{2}{\cal I}_1[q(\cdot) ]- \overline{c} {\cal I}_2 [q(\cdot),\widehat{q}(\cdot)]
+{\cal I}_3[\widehat{q}(\cdot),\lambda] +\lambda\right \}, 
\label{RSfree}
\end{eqnarray}
where
\begin{eqnarray}
{\cal I}_1[q(\cdot) ]&\equiv&\int dA_1 dH_1 q(A_1,H_1) \int dA_2 dH_2 q(A_2,H_2) \cr
&\phantom{=}& \times 
\left [
\left ( \frac{A_2 H_1^2+2 J H_1 H_2 + A_1 H_2^2}{A_1 A_2-1}-\frac{H_1^2}{A_1}-\frac{H_2^2}{A_2} \right ) \right ]_J, 
\label{I1}
\end{eqnarray}
\begin{eqnarray}
{\cal I}_2[q(\cdot),\widehat{q}(\cdot) ]\equiv\int dA dH q(A,H) \int d\widehat{A} d\widehat{H} \widehat{q}(\widehat{A},\widehat{H})
\left (\frac{(H+\widehat{H})^2}{A-\widehat{A}}-\frac{H^2}{A} \right ), 
\label{I2}
\end{eqnarray}
and 
\begin{eqnarray}
{\cal I}_3[\widehat{q}(\cdot),\lambda ] &\equiv& p_1 \int 
\prod_{\mu=1}^{c_1} 
d\widehat{A}_\mu d\widehat{H}_\mu \widehat{q}(\widehat{A}_\mu,\widehat{H}_\mu)
\frac{\left (\sum_{\mu=1}^{c_1} \widehat{H}_\mu \right )^2}{
\lambda-\sum_{\mu=1}^{c_1} \widehat{A}_\mu} \cr
&\phantom{\equiv}&
+p_2 \int 
\prod_{\mu=1}^{c_2} 
d\widehat{A}_\mu d\widehat{H}_\mu \widehat{q}(\widehat{A}_\mu,\widehat{H}_\mu)
\frac{ \left (\sum_{\mu=1}^{c_2} \widehat{H}_\mu \right )^2}{
\lambda-\sum_{\mu=1}^{c_2} \widehat{A}_\mu}. 
\label{I3}
\end{eqnarray}
Here, $[\cdots]_J$ denotes the average with respect to (\ref{biased_dist}). Hereafter, we shall not distinguish between $\Lambda$ and $\left [\Lambda \right]_{\bJ}$ because $\Lambda \to \left [\Lambda \right]_{\bJ}$ should hold with a probability of unity for $N \to \infty$ because of the self-averaging property. The variational functions $q(A,H)$ and $\widehat{q}(\widehat{A},\widehat{H})$ are  joint distributions that come from the RS saddle point calculation, whereas $\lambda$ originates from the constraint of the $\delta$-function in (\ref{eq:partition}). The notation $\mathop{\rm extr}_{X} \{f(X) \}$ generally stands for extremization of $f(X)$ with respect to $X$. A derivation of (\ref{RSfree})--(\ref{I3}) is shown in \ref{RSfree_derivation}. 

\subsection{Cavity method}
An alternative approach, termed the {\em cavity method} \cite{cavity}, is of utility for understanding the physical implications of the seemingly artificial extremization variables $q(A,H)$, $\widehat{q}(\widehat{A},\widehat{H})$ and $\lambda$. In the spirit of mean field theory, directly approximating the multivariate optimization problem of (\ref{eq:quadopt}) by a bunch of single-variable problems as 
\begin{eqnarray}
\mathop{\rm max}_{v_i} \left \{-A_i v_i^2+2 H_i v_i \right \}
\label{singlebody}
\end{eqnarray}
$(i=1,2,\ldots,N)$ is another promising scheme for computing $\Lambda$, wherein the coefficients $A_i$ and $H_i$ are to be determined in a self-consistent manner. In the {cavity method}, this is done by determining the {\em cavity fields} $A_{i \to j}$ and $H_{i \to j}$, which denote the coefficients of (\ref{singlebody}) for the $j$-cavity system, where a node $j$ of the neighbor of a focused node $i$ is removed, by using the belief propagation algorithm \cite{BP,JPCS2010,BiloriSemerjian2010}:
\begin{eqnarray}
&&\widehat{A}_{i \to j} = \frac{1}{A_{i \to j}}, \quad \widehat{H}_{i \to j}= \frac{J_{ji} H_{i \to j}}{A_{i \to j}}, \label{h_step} \\
&&A_{i \to j} =\lambda-\sum_{k \in \partial i \backslash j} \widehat{A}_{k \to i}, \quad 
H_{i \to j} = \sum_{k \in \partial i \backslash j} \widehat{H}_{k \to i}. \label{v_step}
\end{eqnarray}
Here, $\lambda$ is a Lagrange multiplier for introducing  the constraint $|\bv|^2=N$ of (\ref{eq:quadopt}) while new auxiliary variables $\widehat{A}_{i \to j}$ and $\widehat{H}_{i \to j}$ are sometimes termed the {\em cavity biases}. $\partial i$ denotes the neighbor of $i$ and $\partial i \backslash j$ stands for a set defined by removing node $j$ from $\partial i$. After determining the cavity fields/biases, the coefficients of the approximate objective functions are found to be
\begin{eqnarray}
A_i=\lambda-\sum_{k \in \partial i} \widehat{A}_{k \to i}, \quad 
H_i=\sum_{k \in \partial i} \widehat{H}_{k \to i}. 
\label{full}
\end{eqnarray} 

In a random sparse network, the typical lengths of cycles in the network grow as $O(\ln N)$, which means that the system can be locally regarded as a {\em tree}, ignoring any feedback effects. This allows us to characterize the macroscopic properties of the objective system by utilizing the distributions of the cavity fields/biases $q(A,H)=(\sum_{i=1}^N |\partial i|)^{-1} \sum_{i=1}^N \sum_{j \in \partial i} \delta(A-A_{j \to i})\delta(H-H_{j \to i})$ and $\widehat{q}(\widehat{A},\widehat{H})=(\sum_{i=1}^N |\partial i|)^{-1} \sum_{i=1}^N \sum_{j \in \partial i} \delta(\widehat{A}-\widehat{A}_{i\to j}) \delta(\widehat{H}-\widehat{H}_{i\to j})$, where $|{\cal S}|$ stands for the number of elements in the set ${\cal S}$. Equations (\ref{h_step}) and (\ref{v_step}) indicate that $q(A,H)$ and $\widehat{q}(\widehat{A},\widehat{H})$ are determined in a self-consistent manner:
\begin{eqnarray}
&&\widehat{q}(\widehat{A},\widehat{H})=\int dA dH q(A,H)
\left [\delta\left (\widehat{A}- \frac{1}{A} \right ) \delta \left (\widehat{H}-
\frac{J H}{A} \right ) \right ]_J, 
\label{h_step_dist}\\ 
&& q(A,H)=r_1 \int \prod_{\mu=1}^{c_1-1} 
d\widehat{A}_\mu d \widehat{H}_\mu 
\widehat{q}(\widehat{A}_\mu, \widehat{H}_\mu)
\delta \left (\!A\!-\!\lambda\!+\!\sum_{\mu=1}^{c_1-1} \widehat{A}_\mu \!\right )
\delta \left (\!H\!-\! \sum_{\mu=1}^{c_1-1} \widehat{H}_\mu \!\right ) \cr 
&&\hspace*{.5cm}
+ r_2 \int \prod_{\mu=1}^{c_2-1} 
d\widehat{A}_\mu d \widehat{H}_\mu 
\widehat{q}(\widehat{A}_\mu, \widehat{H}_\mu)
\delta \left (\!A\!-\!\lambda\!+\!\sum_{\mu=1}^{c_2-1} \widehat{A}_\mu \!\right )
\delta \left (\!H\!-\! \sum_{\mu=1}^{c_2-1}\widehat{H}_\mu\! \right ), 
\label{v_step_dist}
\end{eqnarray}
where $r_1\equiv c_1p_1/\overline{c}$ represents the probability that one terminal node has degree $c_1$ when a link is chosen randomly from the connectivity network and similarly for $r_2 \equiv c_2p_2/\overline{c}$. On the other hand, (\ref{full}) means that the joint distribution of $A_i$ and $H_i$ of (\ref{singlebody}) is 
\begin{eqnarray}
&&Q(A,H)=p_1 \int \prod_{\mu=1}^{c_1} 
d\widehat{A}_\mu d \widehat{H}_\mu 
\widehat{q}(\widehat{A}_\mu, \widehat{H}_\mu)
\delta \left (\!A\!-\!\lambda\!+\!\sum_{\mu=1}^{c_1} \widehat{A}_\mu \!\right )
\delta \left (\!H\!-\! \sum_{\mu=1}^{c_1}\widehat{H}_\mu \!\right ) \cr 
&&\hspace*{.5cm}
+ p_2 \int \prod_{\mu=1}^{c_2} 
d\widehat{A}_\mu d \widehat{H}_\mu 
\widehat{q}(\widehat{A}_\mu, \widehat{H}_\mu)
\delta \left (\!A\!-\!\lambda\!+\!\sum_{\mu=1}^{c_2} \widehat{A}_\mu \!\right )
\delta \left (\!H\!-\! \sum_{\mu=1}^{c_2}\widehat{H}_\mu\! \right ), 
\label{full_dist} 
\end{eqnarray}
which leads to the extremization condition with respect to the Lagrange multiplier:
\begin{eqnarray}
1=\int dA dH Q(A,H) \left (\frac{H}{A} \right )^2. 
\label{Lagrange_cond}
\end{eqnarray}

It is noteworthy that (\ref{h_step_dist}), (\ref{v_step_dist}) and (\ref{Lagrange_cond}) exactly constitute the extremization condition of (\ref{RSfree}). This allows us to interpret $q(A,H)$, $\widehat{q}(\widehat{A},\widehat{H})$ and $\lambda$ in (\ref{RSfree}) as distributions of the cavity fields/biases and the Lagrange multiplier, respectively. This interpretation indicates that the supports of $q(A,H)$, $\widehat{q}(\widehat{A},\widehat{H})$ and $Q(A,H)$ cannot be extended to the region of neither $A<0$ nor $\widehat{A}<0$ in order to make the approximate single body maximization problems (\ref{singlebody}) well-posed. This condition plays a crucial role in the later analysis. 

\section{Two solvable examples}
\label{section4}
Now we are ready to tackle the first eigenvalue problem. However, solving the problem exactly is still difficult since it involves functional equations of (\ref{h_step_dist}) and (\ref{v_step_dist}). Therefore, we shall first analyze two solvable examples in order to get insights into constructing appropriate approximation schemes. 

\subsection{Single-degree model}
The first example is the case in which $p_1=1$ exactly holds, which means that all nodes possess the same degree $c_1$. We will refer to this example as the {\em single-degree model}. Equations (\ref{h_step_dist}) and (\ref{v_step_dist}) imply that marginal distributions $q(A)=\int dH q(A,H)$ and $\widehat{q}(\widehat{A})=\int d \widehat{H} \widehat{q}(\widehat{A},\widehat{H})$ generally constitute a set of closed equations while $q(H)=\int dA q(A,H)$ and $\widehat{q}(\widehat{H})=\int d \widehat{A} \widehat{q}(\widehat{A},\widehat{H})$ do not. In particular, in the case of $p_1=1$, for which $r_1=1$ and $r_2=0$ hold, this allows us to assume that the distributions are of the forms $q(A)=\delta(A-a)$ and $\widehat{q}(\widehat{A})=\delta(\widehat{A}-\widehat{a})$. Inserting these into (\ref{RSfree}) yields 
\begin{eqnarray}
\Lambda &=&
\mathop{\rm extr} \left \{
c_1 \left (\frac{am_2+ \Delta m_1^2}{a^2-1} \right ) 
- c_1 \left (\frac{m_2+2m_1 \widehat{m}_1+ \widehat{m}_2}{a-\widehat{a}} \right ) \right . \cr
&& \left . + \frac{c_1(\widehat{m}_2-\widehat{m}_1^2)+c_1^2 \widehat{m}_1^2 }{\lambda-c_1 \widehat{a}} +\lambda \right \}, 
\label{singleRS}
\end{eqnarray}
where $m_1$ and $m_2$ are the first and second moments (about the origin) with respect to $q(H|A)=q(H)$, and similarly for $\widehat{m}_1$ and $\widehat{m}_2$. The extremization is 
carried out 
with respect to all variables except for $\Delta$. After some algebra, the extremization conditions of (\ref{singleRS}) can be summarized as
\begin{eqnarray}
&&\widehat{a}=\frac{1}{\lambda-(c_1-1) \widehat{a}}, \label{EMAcond_1}\\
&&\widehat{m}_1=\frac{\Delta(c_1-1) \widehat{m}_1}{ \lambda-(c_1-1) \widehat{a}}, \label{EMAcond_2}\\
&&
\frac{(c_1-1)(c_1-2)\widehat{m}_1^2}{\left (\lambda-(c_1-1) \widehat{a}\right )^2 }
=\left (1- \frac{c_1-1}{\left (\lambda-(c_1-1)\widehat{a} \right )^2} \right )\widehat{m}_2, \label{EMAcond_3}\\
&&\frac{c_1(\widehat{m}_2-\widehat{m}_1^2)+c_1^2 \widehat{m}_1^2 }{(\lambda-c_1 \widehat{a})^2}=1. \label{EMAcond_4}
\end{eqnarray}
Equation (\ref{EMAcond_2}) indicates that the solutions can be classified into two types depending on whether $\widehat{m}_1$ vanishes or not: 

\begin{itemize} 
\item {$\widehat{m}_1 \ne 0$:} Equation (\ref{EMAcond_2}) means that $\Delta(c_1-1)/(\lambda-(c_1-1)\widehat{a})=1$ holds for $\widehat{m}_1 \ne 0$. This, in conjunction with (\ref{EMAcond_1}), gives
\begin{eqnarray}
\lambda=(c_1-1) \Delta+\frac{1}{\Delta}, 
\label{ferro}
\end{eqnarray}
and $\widehat{a}=1/(\Delta(c_1-1))$. Inserting these values into (\ref{EMAcond_3}) and (\ref{EMAcond_4}) yields nonzero values of $\widehat{m}_1$ and $\widehat{m}_2$, and the positivity of $\widehat{m}_2$ makes this solution valid only for $\Delta > \Delta_{\rm c}=1/\sqrt{c_1-1}$. 

\item $\widehat{m}_1 = 0$: Equation (\ref{EMAcond_3}) means that $(c_1-1)/(\lambda-(c_1-1)\widehat{a})^2=1$ holds for $\widehat{m}_1=0$. This, in conjunction with (\ref{EMAcond_1}), gives
\begin{eqnarray}
\lambda=2 \sqrt{c_1-1}, 
\label{bandedge}
\end{eqnarray}
and $\widehat{a}=1/\sqrt{c_1-1}$. Inserting these and $\widehat{m}_1=0$ into (\ref{EMAcond_4}) yields the value of $\widehat{m}_2$. 
\end{itemize}
In both cases, $\Lambda=\lambda$ after extremization. Therefore, the first eigenvalue of the single-degree model can be written as
\begin{eqnarray}
\Lambda =\left \{
\begin{array}{ll}
(c_1-1) \Delta+1/\Delta, & \ \Delta > \Delta_{\rm c}, \cr
2\sqrt{c_1-1}, & \ \Delta \le \Delta_{\rm c}.
\end{array}
\right .
\label{singleLambda}
\end{eqnarray}

\begin{figure}
\begin{center}
\includegraphics[angle=270,width=16cm]{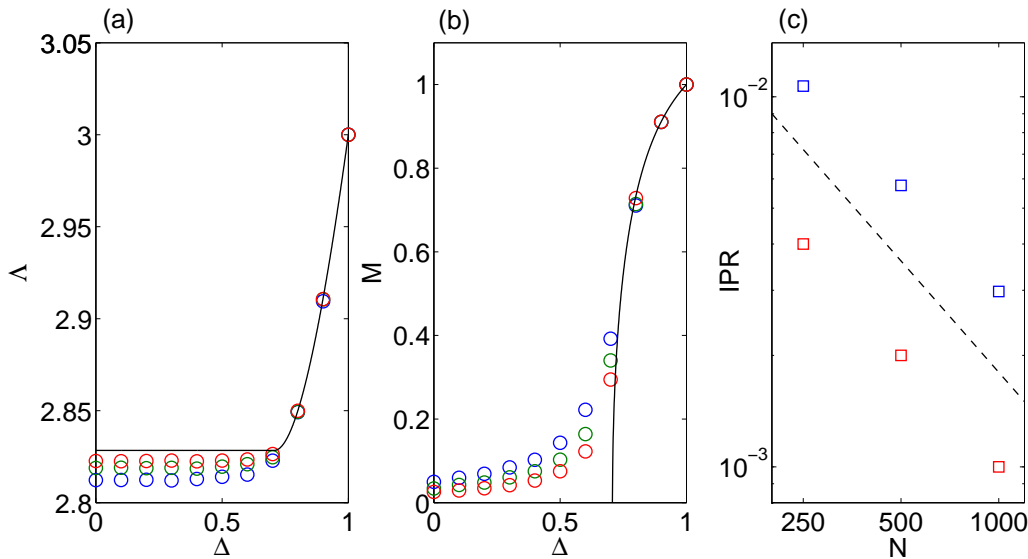}
\caption{(Color online) Theoretical predictions and experimental results for the single-degree model of $c_1=3$. (a) The first eigenvalue. (b) $M=N^{-1}\left |\sum_{i=1}^N V_i \right |$ of the first eigenvector $\bV=(V_i)$. 
Symbols represent averages over 1000 experiments for $N=250$, $500$, and $1000$ systems from the bottom and the top in (a) and (b), respectively. 
The solid curves are the theoretical predictions 
(\ref{singleLambda}) and (\ref{Mdefine}) for (a) and (b), respectively. 
(c) Symbols denote IPR of $\bV$ for $\Delta =0$ and $1$ from the top.
A slope of $O(N^{-1})$ is shown as a broken line for reference.   
}
\label{regular}
\end{center}
\end{figure}


Inserting the functional forms of 
$q(A)=\delta(A-a)$ and $\widehat{q}(\widehat{A})=\delta(\widehat{A}-\widehat{a})$
into (12) and (13) 
yields a self-consistent equation for $q(H)$ as
$q(H)=\int \prod_{\mu=1}^{c_1-1} d H_\mu q(H_\mu) \left [ \delta(H-\sum_{\mu=1}^{c_1-1} J_\mu H_\mu/a ) \right ]_{\{J_\mu\}}$. 
This indicates that the moment generating function of $q(H)$, $g_H(t) \equiv \int dH q(H) e^{tH}$,  
satisfies a relation 
$g_H(t) =\left (\frac{1+\Delta}{2} g_H(t/a)+\frac{1-\Delta}{2} g_H(-t/a) \right )^{c_1-1}$. 
The first and second moments, $m_1$ and $m_2$, of $q(H)$ are determined by 
solving the extremization problem of (16). 
For higher moments of degree $n \ge 3$, taking $n$-th derivative of $g_H(t)$ at $t=0$ offers a formula that 
evaluates $n$-th moment, $m_n$, of $q(H)$ from the moments of lower degrees 
$m_1, m_2, \ldots, m_{n-1}$. 
In particular, the formulae for $m_3$ and $m_4$ are provided as
$
m_3= (a^3 - (c_1-1))^{-1}
(c_1-1)(c_1-2)\left (3 m_2 m_1+ (c_1-3) m_1^3 \right )
$ and 
\begin{eqnarray}
&&m_4=\frac{(c_1-1)(c_1-2)}{a^4-(c_1-1)} \cr
&& \hspace*{1cm}\times
\left (3 m_2^2+4 m_3m_1+ 6 (c_1-3) m_2 m_1^2 +(c_1-3)(c_1-4) m_1^4  \right ), 
\end{eqnarray}
respectively. These guarantee that moments of $q(H)$ are finite at least up to the fourth degree. 
As $n$-th moment of the distribution of entries of the first eigenvector 
$P(V) \equiv N^{-1} \sum_{i=1}^N \left [ \delta(V-V_i) \right ]_{\bJ}
=\int \prod_{\mu=1}^{c_1} dH_\mu q(H_\mu) \left [ \delta (V-\sum_{\mu=1}^{c_1} J_\mu H_\mu /(a(\lambda-c_1/a)) ) \right ]_{\{J_\mu\}}$
can be evaluated from $m_1,m_2,\ldots, m_n$, this 
indicates that the fourth moment of $P(V)$ is finite, and therefore
IPR of the single-degree model vanishes as $O(N^{-1})$ as $N \to \infty$. 

The above computation also implies that the first moment of $P(V)$ is given as
\begin{eqnarray}
M=\left \{
\begin{array}{ll}
c_1(c_1-1)\Delta^2 \widehat{m}_1/((c_1-1)^2\Delta^2-1), & \Delta> \Delta_{\rm c}, \cr
0, & \Delta \le \Delta_{\rm c}. 
\end{array}
\right .
\label{Mdefine}
\end{eqnarray}
This indicates that
$\bV$ for $\Delta > \Delta_{\rm c}$
is macroscopically polarized in the direction of $(1,1,\ldots,1)^{\rm T}$, although the objective function $\bv^{\rm T} \bJ \bv$ is invariant under the transformation of $\bv \to -\bv$. This is analogous to the spontaneous symmetry breaking observed in models of ferromagnetism, and therefore we will term the solutions of this type {\em ferromagnetic solutions}. On the other hand, (\ref{bandedge}) corresponds to the critical condition that equation (\ref{EMAcond_1}), which is a function of $\widehat{a}$, has complex solutions for a given $\lambda$. The complex solutions of $\widehat{a}$ are generally associated with the eigenvalue distribution of $\bJ$ \cite{Kuhn2008}, and (\ref{EMAcond_1}) actually matches the larger band edge of the asymptotic eigenvalue spectrum of the single-degree model. Therefore, solutions of this type will be referred to as {\em band edge solutions}. 

\subsection{Defect model}
\label{defect_models}
Another solvable model can be created by adding only one node of a larger degree 
$c_2 > c_1$ 
to the single-degree model. We refer to the larger degree node as the {\em center}, indexed as $i=0$. 
Let us pay attention to the tree structure rooted at the center $0$. 
In the following, we approximately handle the network as an infinite tree rooted at $0$
since feedback effects are expected to be negligible  in large random sparse networks
as mentioned in section 3.2. 
Equations (\ref{h_step}) and (\ref{v_step}) indicate that for a given $\lambda$, 
all of the $A$-cavity biases heading for the center are given as the smaller solution of (\ref{EMAcond_1}). 
Using this, the second order coefficient of the center node is provided as 
\begin{eqnarray}
A_0=\lambda-c_2 \widehat{a}.
\label{defect_cond}
\end{eqnarray} 
This means that $A_0 \ge 0$ is a required condition for determining the first eigenvalue, yielding 
\begin{eqnarray}
\Lambda=\lambda=\frac{c_2}{\sqrt{c_2-c_1+1}}. 
\label{defect_solution}
\end{eqnarray}
{The expression of (\ref{defect_solution}) was also obtained recently in a mathematically rigorous manner for $\Delta=1$
\cite{Yamaguchi2012}.}
We will term solutions of this type {\em defect solutions} because of their physical implications shown by the following naive analysis \cite{BiroliMonasson1999}.

Since the tree is free of cycles, one can always convert the eigenvalue problem into one for unit nonzero entries of ${J}_{ij}=1$ by making a gauge transformation. 
Let us denote the distance between node $i$ and $0$ as $d$. Due to the spatial symmetry, the entries of the first eigenvector $V_i$ only depend on $d$. Therefore, we will rewrite their values as $V_d$, which allows us to express the eigenvalue equation as
\begin{eqnarray}
\lambda V_0&=&c_2 V_1, \cr
\lambda V_d&=&V_{d-1}+(c_1-1) V_{d+1} \quad (d=1,2,\ldots)
\label{naive_eigen}
\end{eqnarray}
after the gauge transformation. 

The Perron-Frobenius theorem indicates that $V_d$ is of an identical sign for $\forall{d} \ge 0$. 
In addition, the normalization constraint of $|\bV|^2 =N$ requires the boundary condition $\lim_{d \to \infty} (c_1-1)^d V_d^2 <\infty$. 
This choice of solution reproduces the expression of the first eigenvalue (\ref{defect_solution}) and leads to 
\begin{eqnarray}
V_d=(c_2-c_1+1)^{-d/2} V_0. 
\label{defect_vector}
\end{eqnarray}
This indicates that the first eigenvector is localized in the vicinity of the center yielding a finite IPR,
\begin{eqnarray}
{\rm IPR}=\frac{(c_2-2c_1+2)^2(c_2-c_1+2)}{4(c_2-c_1+1)\left ((c_2-c_1+1)^2-c_1+1 \right )}, 
\label{IPR}
\end{eqnarray}
while the first eigenvector for the single-degree model extends over all nodes making IPR vanish as $N \to \infty$ for both the ferromagnetic and band edge solutions. 

\begin{figure}
\begin{center}
\includegraphics[angle=90,width=13cm]{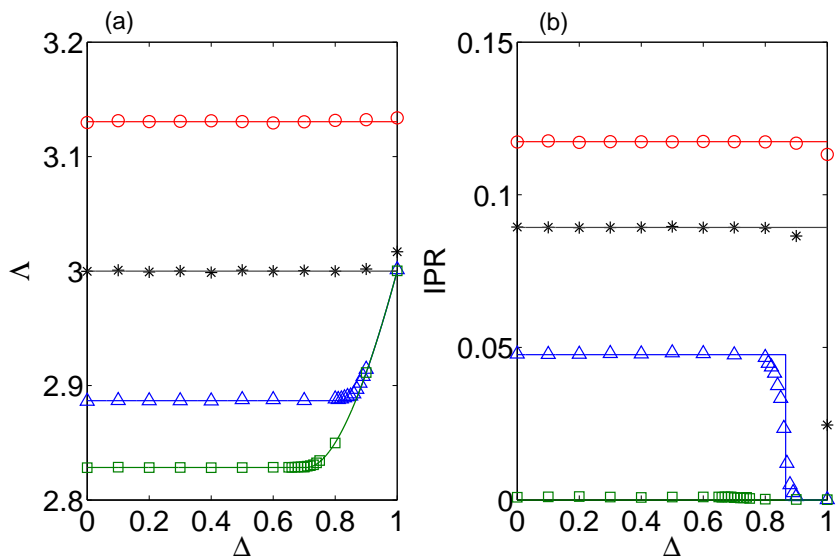}
\caption{(Color online) Theoretical predictions and experimental results for the single-defect models of $c_1=3$ and $c_2=4 \sim 7$. (a) The first eigenvalue. (b) IPR of the first eigenvector. The squares, triangles, asterisks, and circles of both figures represent averages over 100 experiments for $N=8000$ (odd $c_2$) or $8001$ (even $c_2$) systems of $c_2=4,5,6$, and $7$, respectively. The lines are theoretical predictions. }
\label{fig1}
\end{center}
\end{figure}

The analysis shown above means that the existence of a few nodes of larger degree can change the first eigenvalue/eigenvector significantly, as also pointed out in earlier literature \cite{BiroliMonasson1999,SemarjianCugliandolo2002}. Nodes of sufficiently larger degree act as defects receiving more cavity biases than nodes of their surroundings, which boosts the first eigenvalue due to the positivity condition of (\ref{defect_cond}) creating a localized eigenvector. In the current case, this occurs for sufficiently small $\Delta$ if 
$c_2 > 2 (c_1-1)$
and for $\forall{\Delta}\le 1$ if $ c_2 \ge c_1(c_1-1)$. Figure \ref{fig1} compares the theoretical predictions and the results of numerical experiments for (a) the first eigenvalue and (b) IPR of the first eigenvector in the case of $c_1=3$ while varying $c_2$ from $4$ to $7$. 
The numerical experiments were carried out for 
randomly generated networks of finite sizes while the theory is based on the tree approximation. 
In spite of this difference, the theoretical curves of these plots are good matches for the numerical ones. 

In practice, our analysis implies that sufficient precautions must be taken when one utilizes the first eigenvector as a heuristic solution for extracting certain information from a sparse matrix $\bJ$. Even if information on a certain preferential direction is embedded in $\bJ$ as the first eigenvector, the information can be easily hidden by adding only one node of a sufficiently larger degree. To avoid such possibilities, the earlier literature \cite{CS} suggested a preprocessing removing nodes of extraordinarily large degree. 

\begin{figure}
\begin{center}
\includegraphics[angle=90,width=13cm]{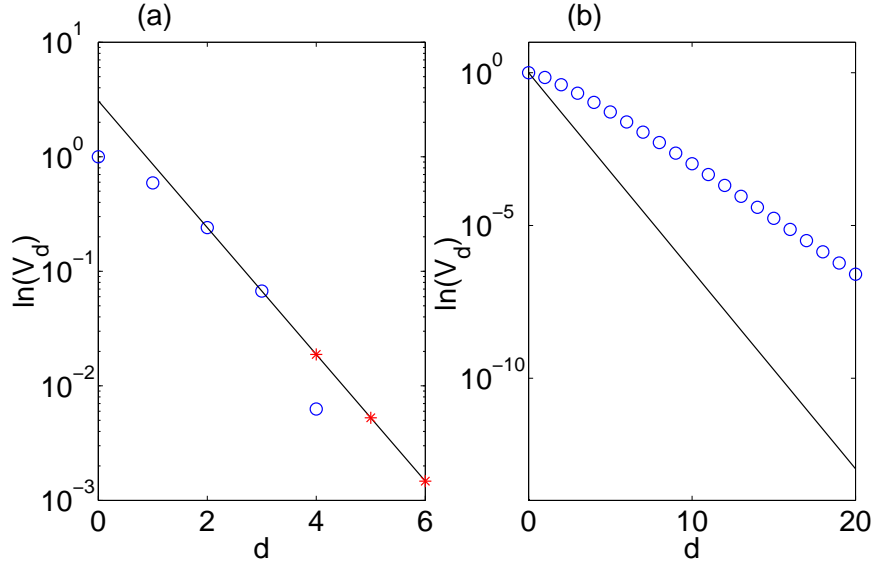}
\caption{(Color online) $\ln (V_d)$ versus $d$ for the recursive equation (\ref{ADgeneral}) for the case of $c_1=3$ and $c_2=7$. 
Circles represent the solution obtained 
from the expression for $1\le d \le g$ of (\ref{ADgeneral}). Lines
stand for the slope of $\ln (\widehat{a}_*(\lambda))$. 
(a): $\lambda $ is chosen so that 
$\ln (V_{g+1})-\ln (V_{g})= \ln (\widehat{a}_*(\lambda))$ holds for $g=2$. 
Asterisks represent the correct solution of (\ref{ADgeneral}) for $d \ge (g+1)+1=4$.
The requirement offers $\lambda = \Lambda=4.1350$.  (b) The case of $\lambda=2 \sqrt{c_2-1}=4.8990$. For $\lambda \ge 2 \sqrt{c_2-1}$, $\ln (V_{d+1} )-\ln (V_d )> \ln (\widehat{a}_*(\lambda))$ holds for $\forall{d} \ge 0$. Therefore, 
there is no $g$ that satisfies (\ref{ADcondition}).}
\label{ADfigure}
\end{center}
\end{figure}

Equation (\ref{defect_cond}) indicates that the eigenvalue of the defect solution becomes larger as the cavity biases coming to the center increase. In addition, the right hand side of (\ref{EMAcond_1}) shows that the cavity biases heading for the center increas as the degrees of surrounding nodes grow. 
This means 
that 
if the number of nodes of the large degree is fixed, the first eigenvalue will be maximized when they are aggregated in the vicinity of the center. As a simple model for representing such situations, let us consider cases in which all nodes within a certain radius $g(=0,1,2,\ldots)$ from the center have the larger degree $c_2$, while the degrees of the other nodes have $c_1$. The case of $g=0$ corresponds to the single-defect model. The analysis above indicates that the first eigenvalue of this aggregated defect model can be estimated by solving the recursive equation,
\begin{eqnarray}
&& V_1=(\lambda/c_2) V_0, \cr
&& V_{d+1}=\left \{
\begin{array}{ll}
 (\lambda V_{d}-V_{d-1} )/(c_2-1) & (d=1,\ldots, g) \cr
(\lambda V_{d}-V_{d-1} )/(c_1-1) & (d=g+1,\ldots)
\end{array}
\right .
\label{ADgeneral}
\end{eqnarray}
under 
the condition that $V_d$ is of an identical sign for $\forall{d} \ge 0$ and $\lim_{d \to \infty} (c_1-1)^d V_d^2 < \infty$.
Given $\lambda$, the solution that satisfies this condition is generally represented as $V_d= const \times \left (\widehat{a}_*(\lambda) \right )^d$ for $d \ge g+1$, where $\widehat{a}_*(\lambda)$ is the smaller solution of (\ref{EMAcond_1}). The condition under which (\ref{ADgeneral}) possesses a solution of this type is expressed as
\begin{eqnarray}
\frac{V_{g+1}}{V_g}=\frac{V_{g+2}}{V_{g+1}}=\widehat{a}_*(\lambda), 
\label{ADcondition}
\end{eqnarray}
which can be used to get the first eigenvalue $\Lambda$ of the aggregated defect model. The IPR of the first eigenvector also comes from (\ref{ADgeneral}). 

Figure \ref{ADfigure} (a) illustrates how to arrived at (\ref{ADcondition}). This figure characterizes the first eigenvalue $\Lambda$ by the condition that the difference $\ln (V_{g+1})-\ln (V_g)$ accords to the target value $\ln (\widehat{a}_*(\lambda))$. For $\lambda \in (2 \sqrt{c_1-1}, 2 \sqrt{c_2-1})$, the left and right terminals of which correspond to the band edge solutions of single-degree models of degree $c_1$ and $c_2$, respectively, the difference $\ln (V_{d+1})-\ln (V_d)$ of the solution of the expression for $1\le d \le g$ of (\ref{ADgeneral}) (circles) can generally vary from a larger value to smaller values than $\ln (\widehat{a}_*(\lambda))$ as $d$ increases from 0. This is because the roots of the characteristic equation of the recursive equation 
are complex numbers, and therefore the solution governed by this recursive equation vanishes in the manner of a damped oscillation as $d$ grows. This means that, for a given $g \ge 0$, there always exists a certain value of $\lambda\in  (2 \sqrt{c_1-1}, 2 \sqrt{c_2-1})$  that satisfies (\ref{ADcondition}). On the other hand, in the region of $\lambda \ge 2 \sqrt{c_2-1}$, the characteristic equation yields roots of positive numbers that are larger than $\ln (\widehat{a}_*(\lambda))$, which means that $\ln (V_{d+1})-\ln (V_d)$ is always larger than $\ln (\widehat{a}_*(\lambda))$. This makes it impossible for (\ref{ADcondition}) to hold (figure \ref{ADfigure} (b)). Consequently, the first eigenvalue of the aggregated defect models increases from the value of (\ref{defect_solution}) to $2\sqrt{c_2-1}$ as $g$ grows from $0$ to $\infty$, while the IPR of the first eigenvector decreases from the value of (\ref{IPR}) to zero. 

{
The convergence behavior of $\Lambda$ is roughly evaluated as follows.
For $\lambda=2 \sqrt{c_2-1}-\epsilon$, where $0 < \epsilon \ll 1$,  
the imaginary part of the roots of the characteristic equation of 
the expression for $1\le d \le g$ of (\ref{ADgeneral}) scales as $O(\epsilon^{1/2})$.  
The radius $g$ that satisfies (\ref{ADcondition}) for given $\lambda$ 
is supposed to be in the same range 
as the period of the damped oscillation caused by the complex roots. This leads to $g \sim O(\epsilon^{-1/2})$, and 
yields an asymptotic relation $\Lambda\sim 2\sqrt{c_2-1} -O(g^{-2})$ for $g \gg 1$ (figure \ref{ADAasympt}).

\begin{figure}
\begin{center}
\includegraphics[angle=90,width=13cm]{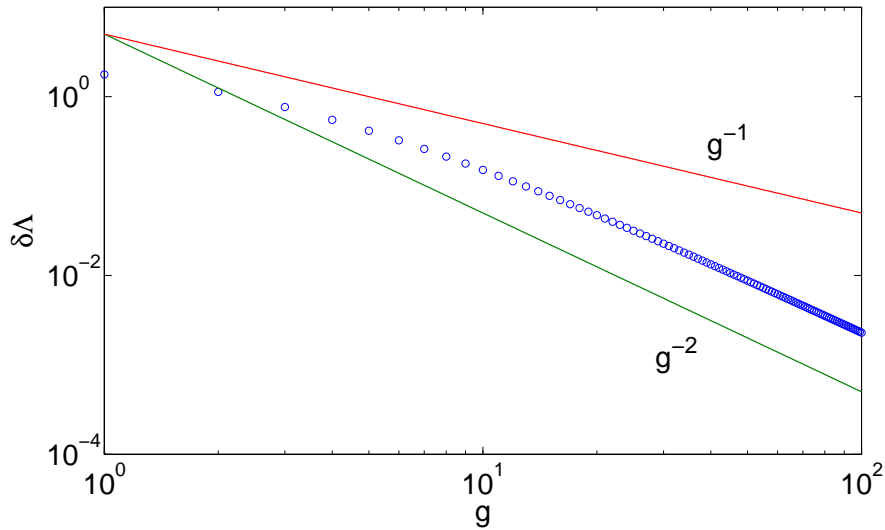}
\caption{(Color online) 
Log-log plot of $\delta \Lambda=2 \sqrt{c_2-1}-\Lambda$ versus $g$ for the solutions of ADA
in the case of $c_1=3$ and $c_2=7$ (symbols). 
Slopes of $g^{-1}$ and $g^{-2}$ (lines) are plotted for reference. 
}
\label{ADAasympt}
\end{center}
\end{figure}
}

\begin{figure}
\begin{center}
\includegraphics[angle=90,width=13cm]{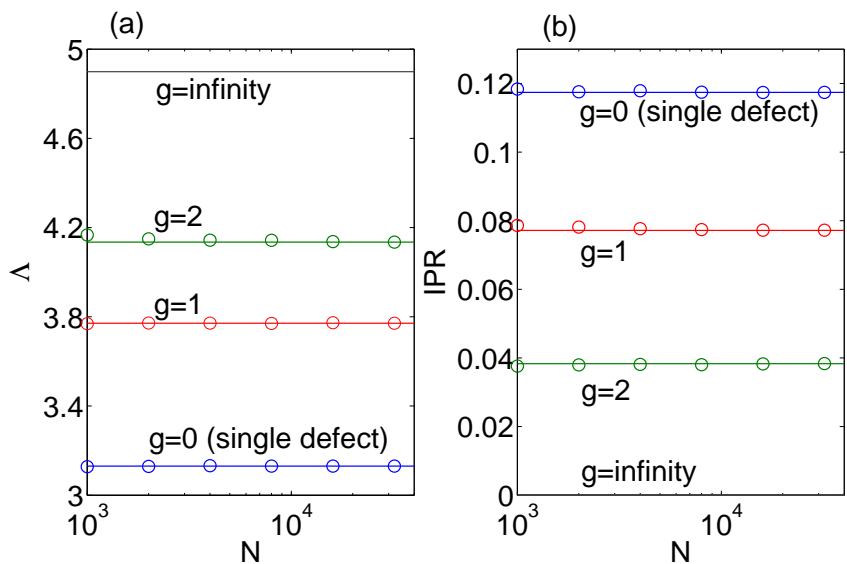}
\caption{(Color online) Theoretical predictions and experimental results for the aggregated defect models of $c_1=3$ and $c_2=7$. (a) The first eigenvalue. (b) IPR of the first eigenvector. Symbols denote averages over 100 experiments for $N=1000 \sim 32000$. Lines represent theoretical predictions.}
\label{ADsummary}
\end{center}
\end{figure}

Figure \ref{ADsummary} shows the first eigenvalue (a) and IPR of the first eigenvector (b) for $c_1=3$ and $c_2=7$ in the case of $\Delta=0$. Experimental results for $N=1000 \sim 32000$ exhibit excellent agreement with the theoretical prediction. 

\section{Approximation for the general case }
\label{section5}
Let us consider a more general situation in which both $p_1$ and $p_2$ are $O(1)$. The framework developed in section \ref{section3} would in principle be valid even in such cases; one would be able to accurately evaluate the typical first eigenvalue by utilizing the solution of (\ref{h_step_dist}) and (\ref{v_step_dist}). Unfortunately, this is difficult in practice. First of all, analytically finding the solution is a hopeless task. Even numerical methods using the standard discretization approach, with the current level of computational resources, have trouble in achieving enough accuracy because of quantization errors. Statistical fluctuations also prevent a sampling approach using population dynamics despite that it performs pretty well in evaluating the bulk profile of the asymptotic eigenvalue spectrum \cite{Kuhn2008,MetzNeriBolle2010}. 

We will avoid such difficulties by taking an alternative strategy. Specifically, we will develop an approximate evaluation scheme that can be handled without solving the functional equations. This scheme does not suffer from either quantization errors or statistical fluctuations, although 
its estimate may be structurally biased. 

\subsection{Effective medium approximation}
The first approximation involves restricting the variational functions in (\ref{RSfree}) to those of the forms $q(A,H)=\delta(A-a)q(H)$ and $\widehat{q}(\widehat{A},\widehat{H})=\delta(\widehat{A}-\widehat{a})\widehat{q}(\widehat{H})$ as assumed in the single-degree model. We call this the {\em effective medium approximation} (EMA) since a similar scheme is referred to by this name in a study of evaluating the asymptotic eigenvalue spectrum \cite{BiroliMonasson1999}. This approximation yields 
\begin{eqnarray}
\Lambda&=&
\mathop{\rm extr} \left \{
\overline{c} \left (\frac{am_2+ \Delta m_1^2}{a^2-1} \right ) 
-\overline{c} \left (\frac{m_2+2m_1 \widehat{m}_1+ \widehat{m}_2}{a-\widehat{a}} \right ) \right . \cr
&& \left . + p_1\frac{c_1(\widehat{m}_2-\widehat{m}_1^2)+c_1^2 \widehat{m}_1^2 }{\lambda-c_1 \widehat{a}} 
+ p_2 \frac{c_2(\widehat{m}_2-\widehat{m}_1^2)+c_2^2 \widehat{m}_1^2 }{\lambda-c_2 \widehat{a}}
+{\lambda} \right \}. 
\label{EMA}
\end{eqnarray}
The implications of the variables are similar to those of (\ref{singleRS}), and the extremization is carried out with respect to all variables except $\Delta$. The extremization condition of (\ref{EMA}) yields the following self-consistent equations:
\begin{eqnarray}
\widehat{a}=\frac{1}{\widehat{a}+\Sigma^{-1}}, 
\label{EMA1}
\end{eqnarray}
\begin{eqnarray}
\widehat{m}_1=\Delta (1-\widehat{a}^2)\left (\frac{r_1 (c_1-1)}{\lambda-c_1\widehat{a}}+\frac{r_2 (c_2-1)}{\lambda-c_2\widehat{a}} \right ) \widehat{m}_1, 
\label{EMA2}
\end{eqnarray}
\begin{eqnarray}
&&\left (\frac{2 \widehat{a}}{1-\widehat{a}^2} 
\left (\frac{r_1(c_1-1)}{\lambda-c_1 \widehat{a} }+\frac{r_2(c_2-1)}{\lambda-c_2 \widehat{a} } \right )
-\left (\frac{r_1(c_1-1)}{(\lambda-c_1 \widehat{a})^2 }+\frac{r_2(c_2-1)}{(\lambda-c_2 \widehat{a})^2 } \right ) \right ) \widehat{m}_1^2 \cr
&&=\left (\frac{r_1 c_1}{(\lambda-c_1\widehat{a})^2}+\frac{r_2 c_2}{(\lambda-c_2\widehat{a})^2}-\frac{\widehat{a}^2+1}{(1-\widehat{a}^2)^2} 
\right ) \widehat{m}_2, 
\label{EMA3}
\end{eqnarray}
\begin{eqnarray}
p_1 \frac{c_1 \widehat{m}_2+c_1(c_1-1) \widehat{m}_1^2 }{(\lambda-c_1 \widehat{a})^2}
+p_2 \frac{c_2 \widehat{m}_2+c_2(c_2-1) \widehat{m}_1^2 }{(\lambda-c_2 \widehat{a})^2}=1, 
\label{EMA4}
\end{eqnarray}
where 
\begin{eqnarray}
\Sigma \equiv \frac{r_1}{\lambda-c_1 \widehat{a}}+\frac{r_2}{\lambda-c_2 \widehat{a}}. 
\label{mean_susceptibility}
\end{eqnarray}
Similarly to the case of the single-degree model, equation (\ref{EMA2}) indicates that the solutions can be classified into two types depending on whether $\widehat{m}_1$ vanishes or not:
\begin{itemize}
\item $\widehat{m}_1\ne 0$: Equation (\ref{EMA2}) means that 
\begin{eqnarray}
\Delta(1-\widehat{a}^2)\left (\frac{r_1(c_1-1)}{\lambda-c_1\widehat{a}}
+\frac{r_2(c_2-1)}{\lambda-c_2\widehat{a}} \right )=1.
\label{EMAferro}
\end{eqnarray}
This and (\ref{EMA1}) together determine $\lambda$ and $\widehat{a}$. Inserting the determined $\lambda$ and $\widehat{a}$ into (\ref{EMA3}) and (\ref{EMA4}) yields $\widehat{m}_1$ and $\widehat{m}_2$. We will refer to this estimate  as the {\em ferromagnetic approximation} (FA). 
\item $\widehat{m}_1 =0$: Equation (\ref{EMA3}) means that 
\begin{eqnarray}
\frac{r_1c_1}{(\lambda-c_1 \widehat{a})^2}+\frac{r_2c_2}{(\lambda-c_2 \widehat{a})^2}
=\frac{(\widehat{a}^2+1)}{(1-\widehat{a}^2)^2}.
\label{EMAsg}
\end{eqnarray}
This and (\ref{EMA1}) determine $\lambda$ and $\widehat{a}$. Equation (\ref{EMAsg}) coincides with the critical condition of $\lambda$ that (\ref{EMA1}) possesses a solution with a complex $\widehat{a}$ (see \ref{complex}), which gives the larger band edge of the asymptotic eigenvalue spectrum under EMA. Therefore, we will call this estimate the {\em band edge approximation} (BEA). Inserting the values of $\lambda$ and $\widehat{a}$ into (\ref{EMA4}) yields $\widehat{m}_2$.
\end{itemize}

\subsection{Aggregated defect approximation}
In addition to the above, the analysis of the defect models offers another criterion for the first eigenvalue. According to the cavity interpretation, $\widehat{a}$ is an exemplary value of the cavity biases $\widehat{A}_{i \to j}$. Therefore, the requirement that (\ref{defect_cond}) must not be negative for any node of the network leads to the condition,
\begin{eqnarray}
\lambda=c_2 \widehat{a}, 
\label{EMAdefect}
\end{eqnarray}
which corresponds to the single-defect approximation (SDA) in the estimate of the eigenvalue spectrum \cite{BiroliMonasson1999,SemarjianCugliandolo2002}. However, this, being combined with (\ref{EMA1}) and (\ref{mean_susceptibility}), always yields a solution of $\widehat{a}=1$ and $\lambda=c_2$, which corresponds to the trivial upper bound of $\Lambda$ for the current bimodal degree model. 

For improving on this result, we can generalize the SDA to higher level of approximation by replacing (\ref{EMAdefect}) with (\ref{ADcondition}) and identifying the solution of (\ref{EMA1}) as $\widehat{a}_*(\lambda)$. We shall refer to the estimate based on this idea as the {\em aggregated defect approximation} (ADA). A similar idea was mentioned in an earlier study on the eigenvalue spectrum \cite{SemarjianCugliandolo2002}. 

Similarly to the argument presented in section \ref{defect_models}, the estimate of the first eigenvalue becomes larger as $g$ grows from $1$ to infinity. In particular, the ADA estimate converges to that of the band edge solution of the single-degree model of $c_2$, $2 \sqrt{c_2-1}$, as $g \to \infty$. Aggregations of the larger degree nodes of arbitrary sizes appear with a probability of unity as $N$ tends to infinity as long as both $p_1$ and $p_2$ are $O(1)$. This indicates that $2 \sqrt{c_2-1}$ is the appropriate estimate of ADA for the current model of $N \to \infty$ irrespective of the details of the degree distribution. 

However, this does not mean that the estimate is practically relevant for explaining the results of experiments on computationally feasible system sizes. The number of nodes of the larger degree $c_2$
surrounding the center of an aggregated defect of radius $g$ is $n_g \equiv c_2+c_2(c_2-1)+c_2(c_2-1)^2+\ldots+c_2(c_2-1)^{g-1}=c_2 ((c_2-1)^{g}-1)/(c_2-2)$. Using this formula, the probability of a node being the center of the aggregated defect is $P_g\simeq p_2 \times r_2^{n_g}$, when both $p_1$ and $p_2$ are $O(1)$. The typical size of the largest aggregation in a network of $N$ nodes can be roughly found using the condition $NP_{g_{\rm max}} \simeq 1$. This yields $n_{g_{\rm max}} \sim O(\ln N)$ and therefore the maximum radius $g_{\rm max}$ typically scales as $O(\ln \ln N)$. 
This, in conjunction with the argument of section \ref{defect_models}, 
indicates that the first eigenvalue behaves as $\Lambda \sim 2 \sqrt{c_2-1}-O((\ln \ln N)^{-2})$. 
The $\ln\ln$-dependence on $N$ implies that $\Lambda$ can be arbitrarily close to $2 \sqrt{c_2-1}$ as $N \to \infty$, but a very large $N$ is necessary for experimentally observing the convergent behavior. 

\subsection{Comparison with experimental results}
\begin{figure}
\begin{center}
\includegraphics[angle=90,width=13cm]{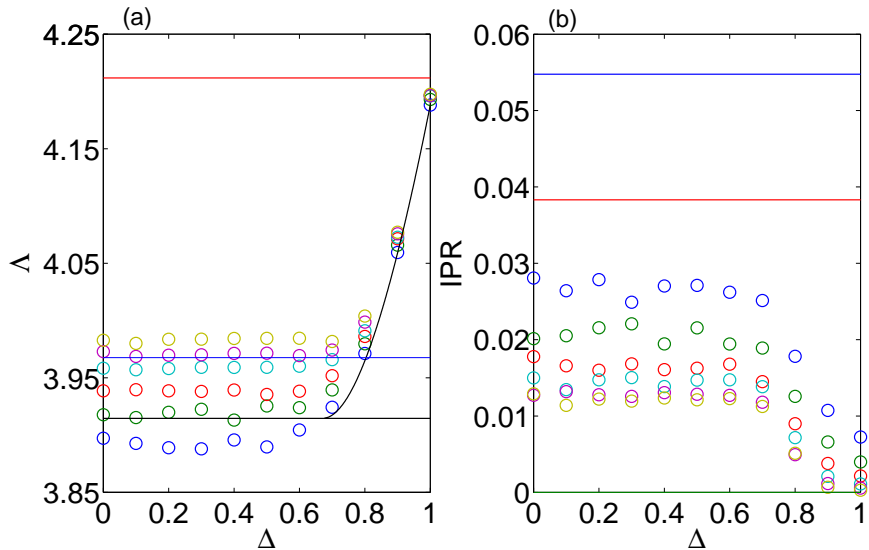}
\caption{(Color online) Theoretical predictions and experimental results in the cases of $c_1=3$, $c_2=7$, and $p_1=1-p_2=0.9$. (a): The first eigenvalue. Symbols represent averages over 100 experiments for $N=1000$, $2000$, $4000$, $8000$, $16000$, and $32000$ systems from the bottom. Lines represent the theoretical predictions by BEA and  ADAs of $g=1$ and $2$ from the bottom while the curve stands for that by FA. The results of ADA indicate that the first eigenvalue converges to $2 \sqrt{c_2-1}=4.8990$ as $N$ tends to infinity. However, due to the $\ln\ln$-dependence of $g_{\rm max}$ on $N$, a very large $N$ would be necessary for experimentally confirming the convergence. (b): IPR of the first eigenvector. Symbols represent averages over 100 experiments for $N=1000$, $2000$, $4000$, $8000$, $16000$, and $32000$ systems from the top. Lines represent the theoretical predictions of ADAs of $g=1$ and $2$ from the top.}
\label{comparison}
\end{center}
\end{figure}

The largest value among the estimates of FA, BEA, and ADA is an approximate estimate of $\Lambda$. To examine the utility of our approximation scheme, we compared the estimated values of $\Lambda$ with the results of numerical experiments for the cases of $c_1=3$, $c_2=7$, and $p_1=1-p_2=0.9$ by varying $N$ from $1000$ to $32000$. The results are depicted in Fig. \ref{comparison} (a). Symbols represent the averages of the first eigenvalues for 100 realizations of matrices. 

As for the choice of parameters, BEA offers an estimate $\Lambda_{\rm BEA} = 3.9146$. As shown in figure \ref{comparison} (a), the estimate of FA, $\Lambda_{\rm FA}$, generally bifurcates from that of BEA at a critical value $\Delta_{\rm c}$, which is evaluated as $0.6762$ for the current parameter choice, as $\Delta$ grows larger from below. The results of the experiments show fairly good accordance with the estimate of FA as $\Delta$ approaches $1$ in the region of $\Delta > \Delta_{\rm c}$. On the other hand, those for $\Delta < \Delta_{\rm c}$ grow gradually as $N$ increases. This is probably because the typical size of the maximum aggregation of the larger degree nodes that dominates the first eigenvalue in the network increases very slowly, as estimated above. ADA estimates $\Lambda_{\rm ADA}$ to be $3.9676$, $4.2119$, and $4.8990$ for $g=1$, $2$, and $\infty$, respectively. The condition of $NP_{g_{\rm max}} \simeq 1$ gives $g_{\rm max} \simeq0.6281 \sim 0.8575 $
for $N=1000 \sim 32000$. This implies that an ADA of $g=1$ is closest to those of the experiments. Actually, it exhibits reasonable consistency with data on larger system sizes $N=8000, 16000$, and $32000$, even though the current estimate of $g_{\rm max}$ is based on a rough argument. 

Figure \ref{comparison} (b) plots the average of IPR for the first eigenvector. The results of the experiments (symbols) are considerably smaller than the theoretical predictions of ADA of $g=1, 2$ (lines). When a network is randomly generated, multiple aggregations of the larger degree nodes appear simultaneously, which reduces the value of IPR. This may be the reason for the significant discrepancy between the theoretical and experimental results. 

\section{Summary}
We investigated the properties of the first (maximum) eigenvalue and its eigenvector (first eigenvector) by using methods of statistical mechanics for sparse symmetric random matrices characterized by a bimodal degree distribution. Employing the replica method, we provided a general formula for evaluating the typical first eigenvalue in the large system size limit. Unfortunately, the replica-based scheme involves functional equations, which are difficult to solve accurately. Therefore, we developed approximate evaluation schemes based on the results for two solvable cases and techniques previously proposed for estimating the eigenvalue spectrum. Our schemes are reasonably consistent with results of experiments when the statistical bias of the positive matrix entries is sufficiently large, and they qualitatively explain why considerably large finite size effects can be observed when the bias is relatively small. 

Promising future research includes an exploration of degree correlated models \cite{Rodgers2010,ShirakiKabashima} as well as a refinement of the approximation schemes. 

\ack
We would like to thank Osamu Watanabe and Koujin Takeda for their helpful discussion. YK acknowledges support by grants from the Japan Society for the Promotion of Science (KAKENHI, No. 22300003) and the Mitsubishi Foundation, and CompView. 

\appendix
\section{A proof of $\Lambda \le c_2$}
\label{LambdaUB}
The Perron-Frobenius theorem guarantees that the inequalities
\begin{eqnarray}
\Lambda&=&\frac{1}{N} \mathop{\rm max}_{\bv} \left \{ \sum_{i, j} J_{ij} v_i v_j \right \} \ \mbox{subj. to } |\bv|^2 =N \cr
&\le& \frac{1}{N} \mathop{\rm max}_{\bv} \left \{ \sum_{i, j} |J_{ij}| |v_i| |v_j| \right \}\ \mbox{subj. to } |\bv|^2 =N \cr
&=& \frac{1}{N} \mathop{\rm max}_{\bv} \left \{ \sum_{i, j} |J_{ij}| v_i v_j \right \} \ \mbox{subj. to } |\bv|^2 =N
\label{quadratic}
\end{eqnarray}
hold for an arbitrary symmetric matrix $\forall{\bJ}=(J_{ij})$. Therefore, we only have to consider the cases in which all nonzero entries are unity. Given such a sample matrix $\bJ$ for which $Np_1$ nodes have degree $c_1$ while the other $Np_2$ nodes have degree $c_2$, we shall add entries of unity, so as to make all nodes have degree $c_2$ while keeping the matrix symmetric. We denote the resultant matrix $\bJ^\prime=(J_{ij}^\prime)$. We also write the first eigenvector of $\bJ$ as $\bV=(V_i)$, assuming a normalization of $|\bV|^2=N$. The Perron-Frobenius theorem ensures that $\forall{V}_i$ is non-negative as well. This indicates that the inequality 
\begin{eqnarray}
\Lambda &=& \frac{1}{N} \sum_{ij} J_{ij} V_i V_j \le \frac{1}{N} \sum_{ij} J_{ij}^\prime V_i V_j \cr
&\le & \frac{1}{N} \mathop{\rm max}_{\bv} \left \{ \sum_{ij} J_{ij}^\prime v_i v_j \right \} 
\ \mbox{subj. to } |\bv|^2 =N 
\label{extraentries}
\end{eqnarray}
holds since entries of $\bV$, $\bJ$ and $\bJ^\prime$ are all non-negative and the number of nonzero entries of $\bJ^\prime$ is larger than that of $\bJ$. The last expression of (\ref{extraentries}) is maximized by $\bv=(1,1,\ldots,1)^{\rm T}$ for any realization of $\bJ^\prime$, which yields $N^{-1} \sum_{ij} J_{ij}^\prime v_i v_j =c_2$. Therefore, $\Lambda \le c_2$ always holds for our ensemble of random matrices. 

\section{Replica approach to finding the first eigenvalue}
\label{RSfree_derivation}
Although we shall focus on the bimodal degree distribution for simplicity, extending the following calculation to general degree distributions is straightforward. To calculate the moment of the partition function (\ref{eq:partition}), we first express the matrix entries as $J_{ij}=J_{ji}=L_{\left \langle ij \right \rangle} B_{\left \langle ij \right \rangle}$, where $\left \langle ij \right \rangle$ denotes the unordered pair of $i$ and $j$. $L_{\left \langle ij \right \rangle}$ is set to unity if there is a link for $\left \langle ij \right \rangle$, and it vanishes, otherwise, and $B_{\left \langle ij \right \rangle}$ is a binary value sampled from (\ref{biased_dist}). Permutation symmetry in indexing the nodes allows us to choose a joint distribution of $\{L_{\left \langle ij \right \rangle}\} \in \{0,1\}^{N(N-1)/2}$,
\begin{eqnarray}
p_L \left (\{L_{\left \langle ij \right \rangle} \} \right )
&=&{\cal N}^{-1} \prod_{i=1}^{N p_1} \delta\left (
\sum_{j \ne i} L_{ \left \langle ij \right \rangle} -c_1\right )
\prod_{i=Np_1+1}^N 
\delta\left (
\sum_{j \ne i} L_{ \left \langle ij \right \rangle} -c_2\right ) \cr
&=&{\cal N}^{-1} 
\prod_{k=1}^{Np_1} \oint \frac{dZ_kZ_k^{-(c_1+1)}}{2 \pi {\rm i} }
\times
\prod_{l=Np_1+1}^{N} \oint \frac{dZ_lZ_l^{-(c_2+1)}}{2 \pi {\rm i} } \cr
&\phantom{=}&
\times
\prod_{i=1}^N Z_i^{\sum_{j \ne i} L_{\left \langle ij \right \rangle} }
\label{eq:link_dist}
\end{eqnarray}
reflecting our assumptions on the graph generation. Here, ${\cal N}$ denotes a constant to normalize $p_L \left (\{L_{\left \langle ij \right \rangle} \} \right )$, ${\rm i}=\sqrt{-1}$, and we have utilized a contour integral expression $\delta(x)=\oint dZZ^{-(x+1)}/(2 \pi {\rm i}) $ for the integer $x$. The joint distribution of $\{B_{\left \langle ij \right \rangle}\} \in \{+1,-1\}^{N(N-1)/2}$ is $p_B\left (\{B_{\left \langle ij \right \rangle}\} \right )=\prod_{\left \langle ij \right \rangle} p_J(B_{\left \langle ij \right \rangle}|\Delta)$ by definition. 

Next, we evaluate the average of $Z^n(\beta;\bJ)$ with respect to these distributions by utilizing an identity $Z^n(\beta;\bJ)=\int \left (\prod_{a=1}^nd\bv^a \delta(|\bv^a|^2-N) \right ) \times \exp \left ((\beta/2) \sum_{a=1}^n (\bv^a)^{\rm T} \bJ \bv^a \right ) =\int \left (\prod_{a=1}^n d\bv^a \delta(|\bv^a|^2-N) \right ) \times \exp \left (\sum_{\left \langle ij \right \rangle } \sum_{a=1}^n \beta L_{\left \langle ij \right \rangle } B_{\left \langle ij \right \rangle } v_i^a v_j^a/2 \right )$. This identity is mathematically valid only for $n \in \mN$. In this evaluation, the following expression appears: 
\begin{eqnarray}
{\cal G}(n)&=&\sum_{\{L_{\left \langle ij \right \rangle}\}, \{B_{\left \langle ij \right \rangle}\}} 
p_B(\{B_{\left \langle ij \right \rangle}\}) 
\prod_{i=1}^N Z_i^{\sum_{j \ne i} L_{\left \langle ij \right \rangle} }
\exp \left (\sum_{\left \langle ij \right \rangle }
\sum_{a=1}^n 
\frac{\beta L_{\left \langle ij \right \rangle } B_{\left \langle ij \right \rangle }v_i^a v_j^a}{2} \right ) \cr
&=&\prod_{\left \langle ij \right \rangle }
 \left (1+Z_i Z_j \prod_{a=1}^n \sum_{B_{\left \langle ij \right \rangle}=\pm 1}
 p_J(B_{\left \langle ij \right \rangle}|\Delta) \exp \left ( \frac{\beta B_{\left \langle ij \right \rangle}v_i^a v_j^a}{2} \right ) \right ) \cr
&=& \exp \left ( \sum_{\left \langle ij \right \rangle}
\ln \left (1+Z_i Z_j \overline{\exp \left (\sum_{a=1}^n \frac{\beta B v_i^a v_j^a}{2} \right )} \right ) \right ) \cr
&\simeq& \exp \left (\sum_{\left \langle ij \right \rangle} 
Z_i Z_j \overline{\exp \left (\sum_{a=1}^n \frac{\beta B v_i^a v_j^a}{2} \right ) }\right ) \cr
& \simeq &
 \exp \left (\frac{N^2}{2} \int d\bu_1 
 {\cal Q}(\bu_1) 
 \int d\bu_2 {\cal Q}(\bu_2)\overline{\exp \left ( \sum_{a=1}^n \frac{\beta B u_1^a u_2^a}{2} \right ) }
 \right ), 
\end{eqnarray}
where $\overline{\exp (\beta B u_1 u_2/2) } \equiv \sum_{B=\pm 1} p_J(B|\Delta)\exp (\beta B u_1 u_2/2)$, $\bu_k \equiv (u_k^1,u_k^2,\ldots,u_k^n)$ $(k=1,2)$, and we have introduced an order parameter function, 
\begin{eqnarray}
{\cal Q}(\bu)\equiv \frac{1}{N} \sum_{i=1}^N Z_i \prod_{a=1}^n \delta(v_i^a-u^a). 
\label{order_param}
\end{eqnarray} 
We shall also introduce a conjugate function $\widehat{Q}(\bu)$ for utilizing an identity for $\forall{\bu}$
\begin{eqnarray}
&&1= \int d {\cal Q}(\bu) \delta \left (\frac{1}{N} \sum_{i=1}^N Z_i \prod_{a=1}^n \delta(v_i^a\!- \!u^a)-{\cal Q}(\bu)
\right ) \cr
&&= \int \frac{N d {\cal Q}(\bu) d \widehat{\cal Q}(\bu)}{2 \pi}
\exp \left (\!\widehat{\cal Q}(\bu) \left (\!\sum_{i=1}^N Z_i \prod_{a=1}^n \delta(v_i^a\! - \! u^a) -N{\cal Q}(\bu)\! \right ) \!\right ), 
\end{eqnarray}
and employ another identity
\begin{eqnarray}
\delta\left (|\bv^a|^2 -N \right )=\int \frac{\beta d\lambda^a}{4 \pi}
\exp \left (-\frac{\beta \lambda^a}{2} \left (\sum_{i=1}^n (v_i^a)^2 -N \right ) \right ). 
\end{eqnarray}
These, in conjunction with employment of the saddle point method for the integration with respect to $\cal Q(\bu)$, $\widehat{\cal Q}(\bu)$, and $\lambda^a$ ($a=1,2,\ldots,n$), lead us to an expression for the average of $Z^n(\beta;\bJ)$:
\begin{eqnarray}
\frac{1}{N} \ln \left [Z^n(\beta;\bJ) \right ]_{\bJ}
&=&\mathop{\rm extr}_{{\cal Q}(\cdot),\widehat{\cal Q}(\cdot),\{\lambda^a\}}
\left \{
\frac{1}{N} \ln {\cal G}(n)- \int d\bu {\cal Q}(\bu) \widehat{\cal Q}(\bu) \right . \cr
& +& p_1 \ln \left (
\int d\bu \exp \left (-\sum_{a=1}^n \frac{\beta \lambda^a (u^a)^2}{2} \right ) 
\widehat{\cal Q}^{c_1} (\bu) \right ) \cr
&+& p_2 \ln \left (
\int d\bu \exp \left (-\sum_{a=1}^n \frac{\beta \lambda^a (u^a)^2}{2} \right ) 
\widehat{\cal Q}^{c_2} (\bu) \right ) \cr
&-&\left . \frac{1}{N} \ln {\cal N} + \sum_{a=1}^n \frac{\beta \lambda^a}{2} \right \}
\label{Replicainteger}
\end{eqnarray}
for $n \in \mN$. 

In the calculation of (\ref{Replicainteger}), we assume that the saddle point is dominated by functions of the form 
\begin{eqnarray}
{\cal Q}(\bu)=T \int dA dH q(A,H) \left (\frac{\beta A}{2 \pi} \right )^{n/2}
\exp\left (-\frac{\beta A}{2} \sum_{a=1}^n \left (u^a-\frac{H}{A} \right )^2 \right ), 
\label{RSQ}
\end{eqnarray}
and 
\begin{eqnarray}
\widehat{\cal Q}(\bu)=\widehat{T} \int d\widehat{A} d\widehat{H} 
\widehat{q}(\widehat{A},\widehat{H}) 
\exp\left (\sum_{a=1}^n \left (\frac{\beta \widehat{A}}{2} (u^a)^2 -\beta \widehat{H} u^a \right )\right ), 
\label{RSQhat}
\end{eqnarray}
where $T$ and $\widehat{T}$ are normalization factors so as to make $q(A,H)$ and $\widehat{q}(\widehat{A},\widehat{H})$ distribution functions. We also assume that $\lambda^a=\lambda$ $(a=1,2,\ldots,n)$ holds at the dominant saddle point. These correspond to the replica symmetric ansatz \cite{replica} in the current system. The saddle point method gives $N^{-1} \ln {\cal N} =(\overline{c}/2)\ln (N \overline{c})-p_1 \ln c_1 !-p_2 \ln c_2 !$. Inserting these into (\ref{Replicainteger}) and extremizing the resultant expression with respect to $T$ and $\widehat{T}$ yields
\begin{eqnarray}
&&\frac{1}{N} \ln \left [Z^n(\beta;\bJ) \right ]_{\bJ} = \mathop{\rm extr}_{q(\cdot),\widehat{q}(\cdot),\lambda}\left \{ 
\frac{\overline{c}}{2} \ln \left ({\cal K}_1 \left [ q(\cdot); n \right ] \right )
- \overline{c} \ln \left ({\cal K}_2 \left [q (\cdot), \widehat{q} (\cdot) ; n \right ] \right ) \right . \cr
&& \hspace*{2cm}\left . +p_1{\cal K}_{31} \ln \left (\left [ \widehat{q}(\cdot),\lambda;n \right ] \right )
+p_2 \ln \left ({\cal K}_{32} \left [ \widehat{q}(\cdot),\lambda;n \right ] \right ) +\frac{n \beta \lambda}{2}\right \} , 
\label{RSgeneral}
\end{eqnarray} 
where 
\begin{eqnarray}
{\cal K}_1 \left [ q(\cdot); n \right ]
&\equiv& \int dA_1 dH_1 q(A_1,H_1) \int dA_2 dH_2 q(A_2,H_2) \cr
&\times & \left [\exp \left (n \beta 
\left (\frac{A_2 H_1^2+2 JH_1 H_2 + A_1H_2^2}{2(A_1 A_2-1)}-\frac{H_1^2}{2A_1}-\frac{H_2^2}{2A_2} \right ) \right ) 
\right ]_J \cr
&\times& \left (\frac{A_1 A_2}{A_1A_2-1} \right )^{n/2},
\label{RSgeneral1}
\end{eqnarray}
\begin{eqnarray}
{\cal K}_2 \left [q (\cdot), \widehat{q} (\cdot) ; n \right ]
&\equiv &
\int dA dH q(A,H) \int d \widehat{A} d\widehat{H}\widehat{q}(\widehat{A},\widehat{H}) \cr
&\times & \exp \left (n \beta \left (
\frac{(H+\widehat{H} )^2}{2(A-\widehat{A})}-\frac{H^2}{2A} \right ) 
\right ) \times \left (\frac{A}{A-\widehat{A}} \right )^{n/2}, 
\label{RSgeneral2}
\end{eqnarray}
\begin{eqnarray}
{\cal K}_{3k} \left [ \widehat{q}(\cdot),\lambda;n \right ]
&\equiv & 
\int \left (\prod_{\mu=1}^{c_k} d\widehat{A}_\mu d \widehat{H}_\mu \widehat{q}(\widehat{A}_\mu, \widehat{H}_\mu) \right ) \times
\exp \left (n \beta \frac{\left (\sum_{\mu=1}^{c_k} \widehat{H}_\mu \right )^2}{\lambda -\sum_{\mu=1}^{c_k} \widehat{A}_\mu } \right ) \cr
&\times & \left (\frac{2 \pi}{\beta \left (\lambda- \sum_{\mu=1}^{c_k} \widehat{A}_\mu \right )} \right )^{n/2},
\ \ (k=1,2).
\label{RSgeneral3}
\end{eqnarray}

Although we estimated $\left [Z^n (\beta;\bJ) \right]_{\bJ}$ for $n \in \mN$ with the saddle point method, the expressions (\ref{RSgeneral})--(\ref{RSgeneral3}) are likely to hold for $n \in \mR$ as well. Therefore, we employ them to evaluate $\left [\Lambda \right]_{\bJ} =2\lim_{\beta \to \infty}(\beta N)^{-1} \left [\ln Z(\beta;\bJ) \right]_{\bJ}=2\lim_{\beta \to \infty} \lim_{n \to 0} (\partial/\partial n) (\beta N)^{-1} \ln \left [Z^n(\beta;\bJ) \right]_{\bJ}$, which yields (\ref{RSfree})--(\ref{I3}). 

\section{Critical condition on emergence of complex solution for (\ref{EMA1})}
\label{complex}
Let us consider a linear perturbation $\widehat{a} \to \widehat{a}+{\rm i} \delta \widehat{a}$ around a fixed point of (\ref{EMA1}) for a given $\lambda$, which yields
\begin{eqnarray}
\delta \widehat{a}=-\frac{1}{(\widehat{a}+\Sigma^{-1})^2} \left (\delta \widehat{a}-\frac{1}{\Sigma^2} \frac{\partial \Sigma}{\partial \widehat{a}} \delta \widehat{a} \right )=\frac{1}{ (\Sigma \widehat{a} +1)^2 } \left (\frac{\partial \Sigma}{\partial \widehat{a}}-\Sigma^2 \right ) \delta \widehat{a}. 
\end{eqnarray}
This means that a critical condition so that (\ref{EMA}) possesses a complex solution is provided as 
\begin{eqnarray}
1=\frac{1}{(\Sigma \widehat{a} +1)^2 } \left (\frac{\partial \Sigma}{\partial \widehat{a}}-\Sigma^2 \right ).
\label{complex_critical}
\end{eqnarray}
Equation (\ref{mean_susceptibility}) indicates that
\begin{eqnarray}
\frac{\partial \Sigma}{\partial \widehat{a}}=\frac{r_1 c_1}{(\lambda- c_1 \widehat{a})^2}+\frac{r_2 c_2}{(\lambda- c_2 \widehat{a})^2}
\end{eqnarray}
and 
\begin{eqnarray}
\Sigma=\frac{\widehat{a}}{1-\widehat{a}^2}
\end{eqnarray}
hold. Inserting these into (\ref{complex_critical}) results in an expression equivalent to (\ref{EMAsg}).

\end{document}